\documentclass[aps,prl,twocolumn,superscriptaddress,longbibliography]{revtex4-1}
\usepackage[usenames, dvipsnames]{color}
\usepackage{graphicx}
\usepackage{color}
\usepackage{epstopdf}
\usepackage{epsfig}
\usepackage{bm}
\usepackage{xcolor}
\usepackage{amsmath}
\usepackage{amsfonts}
\usepackage{float}
\usepackage{subfigure}
\usepackage{verbatim}
\usepackage[version=3]{mhchem}
\usepackage[utf8]{inputenc}
\usepackage{epstopdf}
\usepackage{scalerel}
\usepackage{multirow}
\usepackage{booktabs}
\usepackage{hyperref}
\usepackage[para,flushleft]{threeparttable}

\setcitestyle{super}

\newcommand{\degree}{^{\circ} }

\newcommand{\rb}{\mathbf{r}}

\newcommand{\avg}[1]{\left<#1\right>}
\newcommand{\len}[1]{\left|#1\right|}

\newcommand{\para}[1]{\left(#1\right)}

\newcommand{\kT}{\ensuremath{k_{\rm B}T}}
\newcommand{\Mb}{\ensuremath{\mathcal{M}}}

\newcommand{\vb}{\ensuremath{\mathbf{v}}}

\begin{document}

%\preprint{}

\title{Electronic Paddlewheels Impact the Dynamics of Superionic Conduction in AgI}



\author{Harender S. Dhattarwal}
\author{Richard C. Remsing}
\email[]{rick.remsing@rutgers.edu}
\affiliation{Department of Chemistry and Chemical Biology, Rutgers University, Piscataway, NJ, 08854, USA}

%\thanks{}


%\date{\today}

\begin{abstract}
Solid-state ion conductors hold promise as next generation battery materials.
To realize their full potential, an understanding of atomic-scale ion conduction mechanisms is needed,
including ionic and electronic degrees of freedom. 
Molecular simulations can create such an understanding, however, including a description of electronic structure necessitates computationally expensive methods that limit their application to small scales.
We examine an alternative approach, in which neural network models are used to efficiently sample ionic configurations and dynamics at ab initio accuracy.
Then, these configurations are used to determine electronic properties in a post-processing step. 
We demonstrate this approach by modeling the superionic phase of AgI, in which cation diffusion is coupled to rotational motion of local electron density on the surrounding iodide ions, termed electronic paddlewheels.
The neural network potential can capture the many-body effects of electronic paddlewheels on ionic dynamics, but classical force field models cannot.
Through an analysis rooted the generalized Langevin equation framework, we find that electronic paddlewheels have a significant impact on the time-dependent friction experienced by a mobile cation. 
Our approach will enable investigations of electronic fluctuations in materials on large length and time scales, and ultimately the control of ion dynamics through electronic paddlewheels.
\end{abstract}

%\keywords{}

\maketitle

\raggedbottom

\section{Introduction}
Increasing world energy demands requires increasing energy availability in a renewable and sustainable manner to avoid detrimental environmental effects~\cite{sofian2024energy}.
Renewable energy sources are often intermittent, creating a need for efficient energy storage systems like battery technologies~\cite{Adeyinka2024,Shen2025}.
Most currently available battery technologies rely on liquid electrolytes, typically with Li$^+$ as the relevant electrolyte cation. 
These conventional liquid electrolytes exhibit several problems, including flammability, leakage, and limited electrochemical stability windows~\cite{Thackeray2012LiLimitations,Tarascon2001limitations}. 
Alternatives such as water-in-salt electrolytes~\cite{Suo2015Wise,Suo2016Wibe,Dhattarwal2021Wise}, ionic liquids~\cite{Watanabe2017ILs}, and their hybrids~\cite{Becker2021Hybrid,Dhattarwal2024hybrid} are being developed to overcome some of these limitations. 
Solid-state ion conductors can mitigate many of these issues within solid-state batteries~\cite{Liang2021,Bachman2016}.
However, because of the strong interactions within purely ionic solids, solid-state ion conductors have their own limitations, such as slow ion conduction at room temperature. 
It is therefore advantageous to design new solid-state ion conductors with higher conductivity at intended operating conditions,
and the knowledge-based design of these materials requires a detailed, atomic-scale understanding of ion conduction mechanisms.
We recently demonstrated the existence of electronic paddlewheels in solid-state ion conduction within AgI~\cite{Dhattarwal2024paddle}. 
In the electronic paddlewheel mechanism, the translational motion of one ion type (here Ag$^+$) is coupled to rotations of the electron density on the other ions (here I$^-$).
This is akin to the paddlewheel mechanism proposed for molecular solid-state ion conductors like LiSO$_4$, except electronic degrees of freedom are orientationally disordered instead of a molecule~\cite{Lunden1988LiSO4,zhang2022exploiting}. 
The electronic-structure based perspective on ion conduction provided by electronic paddlewheels opens new avenues for developing design principles, but several fundamental questions need to be addressed before the electronic paddlewheel perspective can be used for materials design. 
Here, we address two of these questions:
(i) The ab initio simulations required to investigate ion conduction are prohibitively expensive and limited to small system sizes and short timescales, so how can these simulations be scaled up? 
(ii) How do electronic paddlewheels impact ionic dynamics? 
To address the first question of scalability, we consider
machine learning potentials (MLPs) as efficient alternatives to ab initio simulations, but these approaches typically lack an explicit description of electronic structure~\cite{Behler2007,Zhang2018,Batzner2022nequip,Thakur2024nequip,Riniker2018,Brooks2021}.
However, within the Born-Oppenheimer approximation, the electronic structure and its fluctuations are determined by the instantaneous configuration of the nuclei (though the forces on the nuclei and the resulting trajectory depends on both nuclear and electronic degrees of freedom). 
Therefore, one may attempt to use MLPs to efficiently access long time scales and even large system sizes, and then use these configurations to examine electronic structure fluctuations. 
Here, we assess the ability of trajectories generated by a machine learning potential to describe the static and dynamic electronic fluctuations responsible for electronic paddlewheels in $\alpha$-AgI. 
We find near perfect agreement between the ab initio and MLP results, because of the ability of the MLP to capture the complex many-body effects arising from electronic paddlewheels.
We contrast the MLP results by also studying a classical, empirical force field (FF) model, which, despite exhibiting  structural averages similar to the ab initio simulations, cannot capture the effects from electronic paddlewheels.

To demonstrate the inability of the FF model to capture electronic paddlewheel effects on ion dynamics, we address the second question regarding how electronic paddlewheels impact ion dynamics. 
Through an analysis of ionic dynamics within the framework of the generalized Langevin equation,
we demonstrate that the classical FF model cannot capture the impact of electronic paddlewheels on ion dynamics.
In particular, we show that electronic paddlewheels within the ab initio and MLP simulations lead to higher frequency force fluctuations than found in the classical FF model, and these additional fluctuations significantly alter the time-dependent friction acting on a silver cation.
This electronic paddlewheel-induced change in friction enhances cation diffusion.

\section{Methods}

We compared the structure and dynamics of AgI from density functional theory based molecular dynamics (DFT-MD) simulations with classical molecular dynamics simulations performed using the Neural Equivariant Interatomic Potential~\cite{Batzner2022nequip} (NequIP) MLP and the Vashishta-Rahman empirical pair potential~\cite{AgI_Vashishta}. 
DFT-MD trajectories were taken from our previous work~\cite{Dhattarwal2024paddle}.
We then selected configurations from trajectories at different temperatures to map atom positions and chemical species to the total potential energy and interatomic forces. 
These configurations correspond to a cubic simulation cell of $\alpha$-AgI with a length of 15.2565~\AA~and 108 atoms~\cite{hull2004superionics}.
We used a total of 3600 configurations (500 at 100~K, 1050 at 300~K, 1050 at 750~K, 500 at 1000~K, and another 500 at 1600~K), out of which 2000 configurations were used for training, 1000 for validation, and 600 for testing the model, all sampled uniformly from the full data set. 
The locality for the features was imposed with a radial cutoff of 6~\AA. 
We used an initial learning rate of 0.005 and a batch size of 5 for the training. 
The learning rate gets rescaled by a factor of 0.5 if the validation loss of atomic forces has not improved over 100 consecutive epochs. 
The training was terminated after around 2100 epochs, once the errors in the forces and energies were sufficiently small;
we used energy and force convergence criteria of $8\times10^{-4}$~eV/atom $9\times10^{-3}$~eV/\AA, respectively. 
Correlation plots between the MLP with DFT energies and forces are shown in Fig.~S1 of the SI and validate the accuracy of the MLP model.
Moreover, the good agreement between the structural and dynamic properties computed from the MLP and DFT simulations indicates that the MLP is able to capture the relevant physics of the system.
We used the Vashishta-Rahman potential for the classical force field-based MD simulations of $\alpha$-AgI,
and we refer the reader to the original work for a listing of the various parameters used in the model~\cite{AgI_Vashishta}.
Within this model, the pairwise interatomic interactions between atom types $i$ and $j$ are a sum of four contributions,
\begin{equation}
u_{ij}(r) = u^{\rm rep}_{ij}(r) + u^q_{ij}(r) + u^{\rm disp}(r) + u^{\rm pol}(r).
\end{equation}
The first term, $u^{\rm rep}_{ij}(r)$, is a repulsive, inverse power law potential,
\begin{equation}
u^{\rm rep}_{ij}(r) = \frac{ H_{ij} }{ r^{\eta_{ij}} },
\end{equation}
where $H_{ij}$ describes the strength of this repulsion and $\eta_{ij}=7$ for all combinations of $i$ and $j$.
The next term corresponds to the Coulomb interactions between ions of type $i$ and $j$ with charges $q_i$ and $q_j$,
\begin{equation}
u^q_{ij}(r) = \frac{q_i q_j}{r},
\end{equation}
where the fixed charges on Ag$^+$ and I$^-$ are $0.6e$ and $-0.6e$, respectively. 
The potential also includes a $r^{-6}$ term to capture the asymptotic behavior of dispersion interactions,
\begin{equation}
u^{\rm disp}_{ij}(r) = -\frac{W_{ij}}{r^6},
\end{equation}
where $W_{ij}$ (and $H_{ij}$) are parameterized to ensure the correct lattice distance at the minimum energy; only $W_{\rm II}$ is non-zero.
The final contribution to the interaction potential arises from electronic polarizabilities,
\begin{equation}
u^{\rm pol}_{ij}(r) = - \frac{ \frac{1}{2}\para{\alpha_i q_i^2 + \alpha_j q_j^2}}{r^4},
\end{equation}
where $\alpha_i$ is the electronic polarizability of ion type $i$.
The Vashishta-Rahman model makes the approximation that only the iodide ions have a non-zero electronic polarizability,
such that $u^{\rm pol}_{ij}(r)$ is non-zero only for I-I and Ag-I interactions;
polarization on the silver cations and any feedback between silver and iodide polarization is ignored,
similar to recent models for metal ions in solution~\cite{li2014taking}.
This last term is not standard for typical pair potentials used to describe ionic interactions and, as discussed further below, is crucial for $u_{ij}(r)$ to produce a realistic description of $\alpha$-AgI.
For both NequIP and classical MD simulations, starting structures were equilibrated for at least 1~ns at 750~K in the canonical ensemble (NVT) using the Nos\'{e}-Hoover thermostat\cite{Nose1984,Hoover1985}.
For consistency, we model the same system size and lattice parameters in all simulations (cell length 15.2565~\AA~and 108~atoms).
To assess finite size effects, we performed a set of simulations on three larger system with 2916, 6912, and 13,500 atoms. The relevant results are shown in the SI.
For the calculation of structural and dynamic properties, equations of motion were propagated for 100~ps in the microcanonical ensemble using velocity Verlet integrator with a timestep of 0.5~fs. 
The last 40~ps of these trajectories were used for analysis. 
Maximally localized Wannier functions (MLWFs) were obtained by minimizing their spreads within CP2K~\cite{CP2K2020,MLWFC2012,Berghold2000}.
The centers of maximally localized Wannier fuctions (MLWFCs) were used for analyzing electronic correlations.
We used the DDEC6 method to compute bond orders quantifying the interaction between Ag$^+$ and I$^-$ ions~\cite{DDEC6}.

\section{Results and Discussion}

\subsection{Structural Correlations}

\begin{figure}[!htb]
\begin{center}
\includegraphics[width=0.49\textwidth]{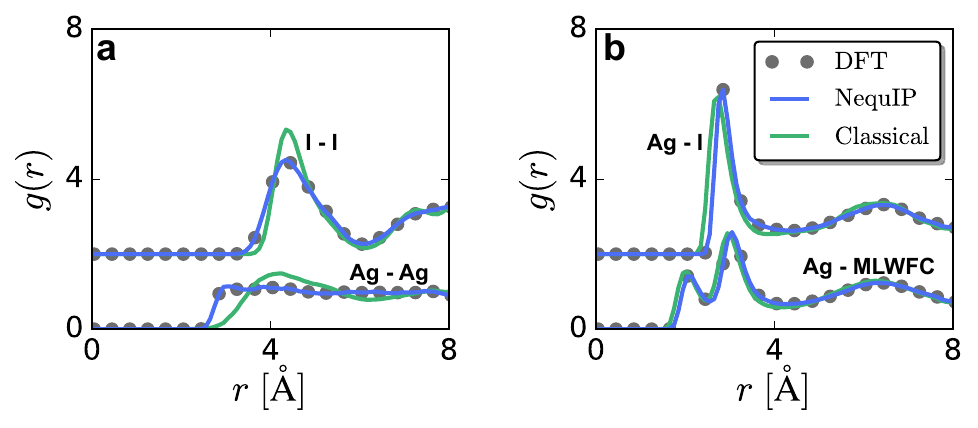}
\end{center}
\caption
{Radial distribution functions, $g(r)$, involving (a) like ion correlations (Ag-Ag and I-I) and (b) unlike ion correlations (Ag-I and Ag-MLWFC). 
The (a) I-I $g(r)$ and (b) Ag-I $g(r)$ are shifted vertically for clarity. 
The Ag-MLWFC $g(r)$ in (b) involves correlations between Ag$^+$ and MLWFCs on I$^-$. }
\label{fig:struct}
\end{figure}

Before quantifying ionic dynamics, we first quantify the pair structure produced by each model through radial distribution functions (RDFs), $g(r)$, Fig.~\ref{fig:struct}.
The cation-cation, anion-anion, and cation-anion RDFs indicate that Ag$^+$ is coordinated by I$^-$ ions and that the cation exhibits liquid-like disordered pair correlations.
Moreover, the RDF for correlations between Ag$^+$ and the I$^-$ MLWFCs indicate that a single MLWFC coordinates a cation while the other three MLWFCs are further away.
All RDFs produced by the NequIP MLP are nearly identical to those of the ab initio simulations (DFT), while the classical force field produces the same qualitative features but a slightly more structured system. 
Surprisingly, the correlations between Ag$^+$ and I$^-$ MLWFCs are described well by both models, suggesting that configurations produced by these models can be used to describe the average structure in the system. 
%

%@@@@@@@@@@@@@@@@@@@@@@@@@@@@@@@@@@@@@@@@@@@@@@
%@@@@@@@@@@@@@@@@@@@@@@@@@@@@@@@@@@@@@@@@@@@@@@

\begin{figure}[!htb]
\begin{center}
\includegraphics[width=0.49\textwidth]{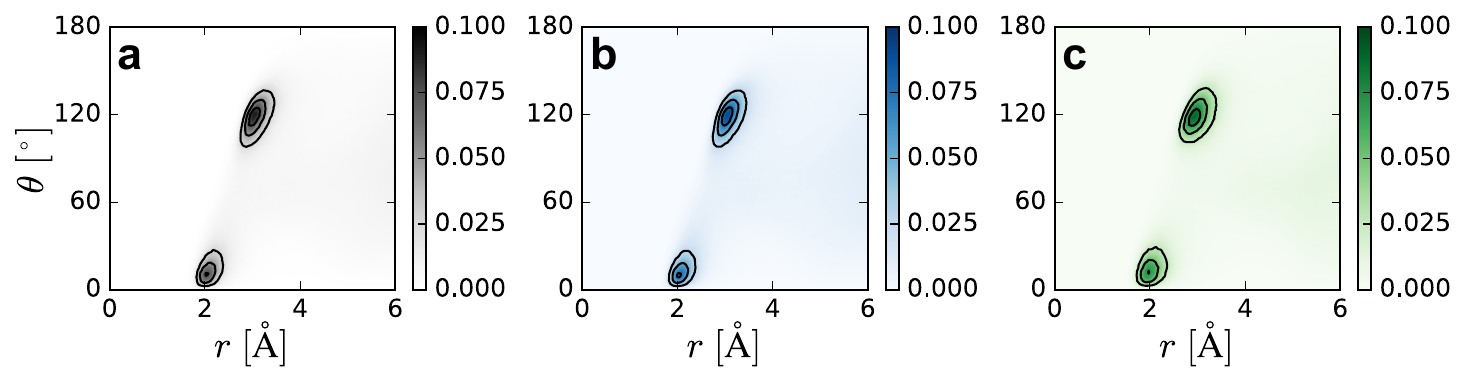}
\end{center}
\caption
{Joint probability distribution of the MLWFC-Ag$^+$ distance ($r$) and the I$^-$-MLWFC-Ag$^+$ angle ($\theta$) from DFT, MLP, and FF models, respectively. 
}
\label{fig:gtheta}
\end{figure}

%%
%
We further examine structural correlations through the joint distribution, $g(r,\theta)$, of the distance, $r$, between Ag$^+$ and I$^-$ MLWFCs and the Ag-MLWFC-I angle, $\theta$, Fig.~\ref{fig:gtheta}.
The joint distribution is consistent with one iodide MLWFC closely and linearly coordinating Ag$^+$ and the other three MLWFCs pointing away from the cation at an angle of 120$\degree$, consistent with the tetrahedral arrangement of MLWFCs on iodide.
The joint distributions are similar for all three models, suggesting that the relevant coordination of Ag$^+$ by iodide MLWFCs is reproduced and both the MLP and the FF capture these three-body correlations at the level of the DFT simulation.
The MLP is trained to capture these correlations by learning the DFT forces as a function of atomic environment. 
It may be surprising that the pairwise Vashishta-Rahman potential is able to capture these complex structural correlations. 
However, this FF includes explicit, though averaged, contributions from atomic polarizabilities via $u^{\rm pol}_{ij}(r)$. 
We suggest that the presence of this polarizability term is critical to producing the correct structure. 
Indeed, removal of this interaction between silver and iodide causes the structure of AgI to become disordered.
%

%@@@@@@@@@@@@@@@@@@@@@@@@@@@@@@@@@@@@@@@@@@@@@@
%@@@@@@@@@@@@@@@@@@@@@@@@@@@@@@@@@@@@@@@@@@@@@@

\subsection{Translational Dynamics}

\begin{figure}[tb]
\begin{center}
\includegraphics[width=0.49\textwidth]{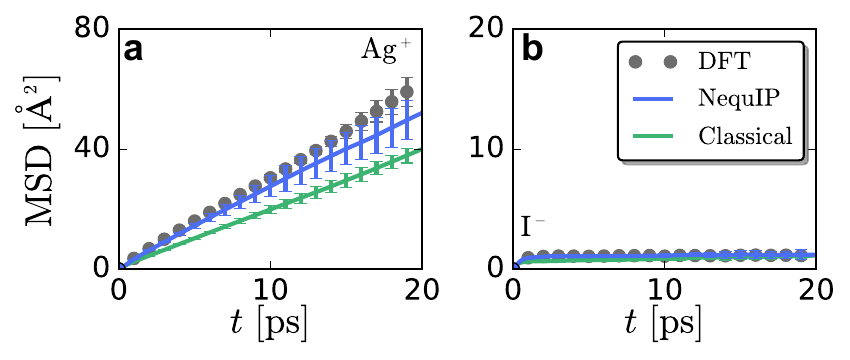}
\end{center}
\caption
{Mean square displacement of (a) Ag$^+$ cations and (b) I$^-$ anions obtained from DFT, MLF, and FF models.
}
\label{fig:msd}
\end{figure}

In the superionic phase, $\alpha$-AgI, Ag$^+$ diffuses, but I$^-$ does not.
This diffusion can be observed in both the mean squared displacement (MSD), $\Delta r^2(t)$, and velocity autocorrelation function (VACF), $C_v(t)$, of each ion type. 
The MSD is defined as
\begin{equation}
\Delta r^2(t) = \avg{\sum_j \len{\rb_j(t)-\rb_j(0)}^2},
\end{equation}
where $\rb_j(t)$ is the position of particle $j$ at time $t$.
The VACF is defined as
\begin{equation}
C_v(t) = \frac{\avg{\sum_j \vb_j(0)\cdot \vb_j(t)}}{\avg{\sum_j \vb_j(0)\cdot \vb_j(0)}},
\end{equation}
where $\vb_j(t)$ is the velocity of particle $j$ at time $t$.
The MSD for Ag$^+$ quickly becomes linear in time, while that for I$^-$ rapidly plateaus to a constant value, indicating that all three models produce a superionic phase
in which cations diffuse and anions do not, Fig.~\ref{fig:msd}.
The slope of the MSD is related to the diffusion coefficient by $6Dt = \lim_{t\rightarrow\infty} {\rm MSD}(t)$.
Our MSDs indicate that the diffusion coefficient is similar for DFT and the MLP, although the MLP diffusion coefficient may be slightly lower but is within the error bars of that from DFT.
The diffusion coefficient for the FF is significantly smaller than the other two systems, resulting from the more ordered structure it produces.
We note that diffusion coefficients and therefore MSDs exhibit significant finite size effects~\cite{celebi2021finite}, and we show that this is also true at least in the FF model in SI Fig.~S3.

\begin{figure}[tb]
\begin{center}
\includegraphics[width=0.49\textwidth]{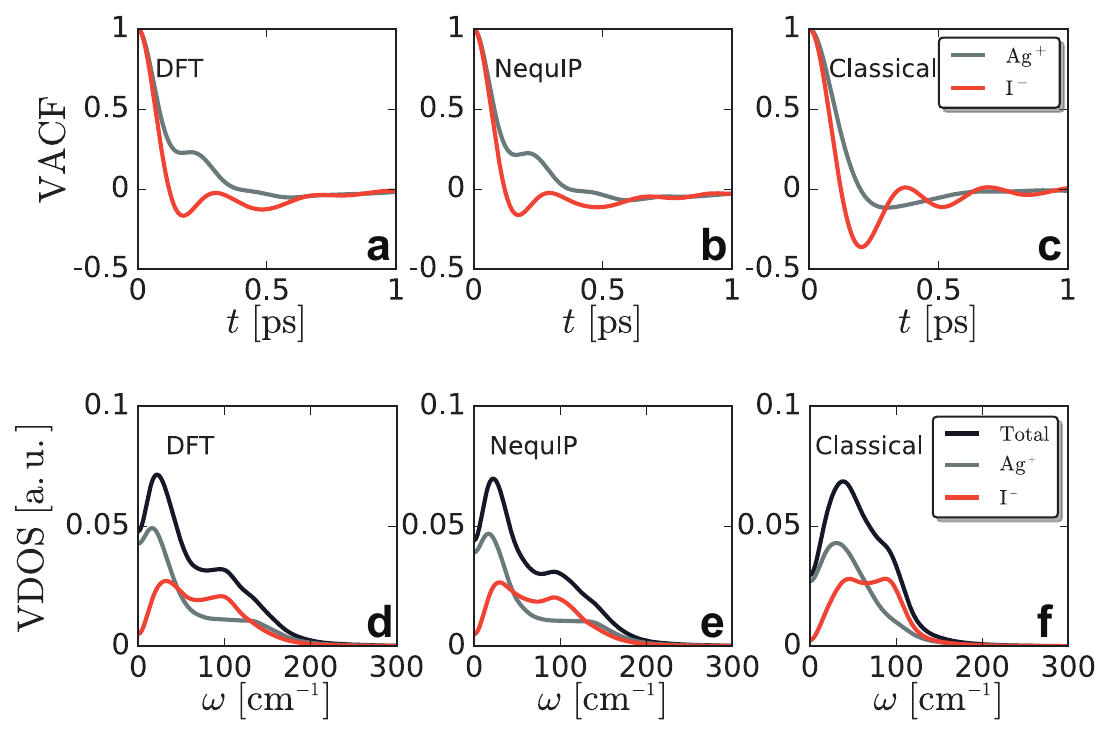}
\end{center}
\caption
{(a-c) Velocity autocorrelation functions (VACFs) and (d-f) correspond vibrational densities of states (VDOS) for the (a,c) DFT, (b,e) MLP, and (c,f) FF models under study. 
}
\label{fig:vacf}
\end{figure}

Similar trends can be observed in the VACFs and their corresponding power spectra or vibrational densities of states (VDOS), $\hat{C}_v(\omega)$, Fig.~\ref{fig:vacf}.
The DFT and MLP VACFs are nearly identical, while that for the FF displays more pronounced negative minima for both ions.
The more pronounced minima indicate that the structure produced by the FF is stiffer, as suggested by RDFs and MSDs.
Moreover, the decay of the silver VACF produced by the FF closely follows that of the iodide, suggesting a strong coupling between the translational dynamics of cations and anions. 
In contrast, silver VACFs produced by DFT and MLP simulations exhibit a positive decay, while the iodide VACFs exhibit a negative decay to zero. 
These differences suggest that their is additional coupling between the cation and anion in the DFT and MLP models, beyond coupling between translations.
The power spectra are also similar for DFT and the MLP, while that for the FF is different, reflecting differences in their dynamics.
All three VDOS indicate diffusion of the cation, because the cation (and total) power spectrum is finite at $\omega=0$,
albeit $\hat{C}_v(0)$ is smaller in the FF model, consistent with its smaller diffusion coefficient.
The VDOS for the anion is zero in the same limit, indicating a lack of diffusion in all three models.
The Ag$^+$ power spectra from DFT and the MLP exhibit a low frequency peak followed by a plateau region that extends to higher frequencies, while the FF VDOS exhibits only a single broad low frequency peak.
The additional high frequency VDOS in the DFT and MLP models is consistent with the presence of additional oscillations and the qualitatively different form of the decay in their VACFs, as compared to that of the FF. 
Similarly, the I$^-$ power spectra are broader from DFT and MLP than that produced by the FF, which reflects stiffening of AgI in the FF.
%

%@@@@@@@@@@@@@@@@@@@@@@@@@@@@@@@@@@@@@@@@@@@@@@
%@@@@@@@@@@@@@@@@@@@@@@@@@@@@@@@@@@@@@@@@@@@@@@

%@@@@@@@@@@@@@@@@@@@@@@@@@@@@@@@@@@@@@@@@@@@@@@
%@@@@@@@@@@@@@@@@@@@@@@@@@@@@@@@@@@@@@@@@@@@@@@

%@@@@@@@@@@@@@@@@@@@@@@@@@@@@@@@@@@@@@@@@@@@@@@
\begin{figure}[tb]
\begin{center}
\includegraphics[width=0.49\textwidth]{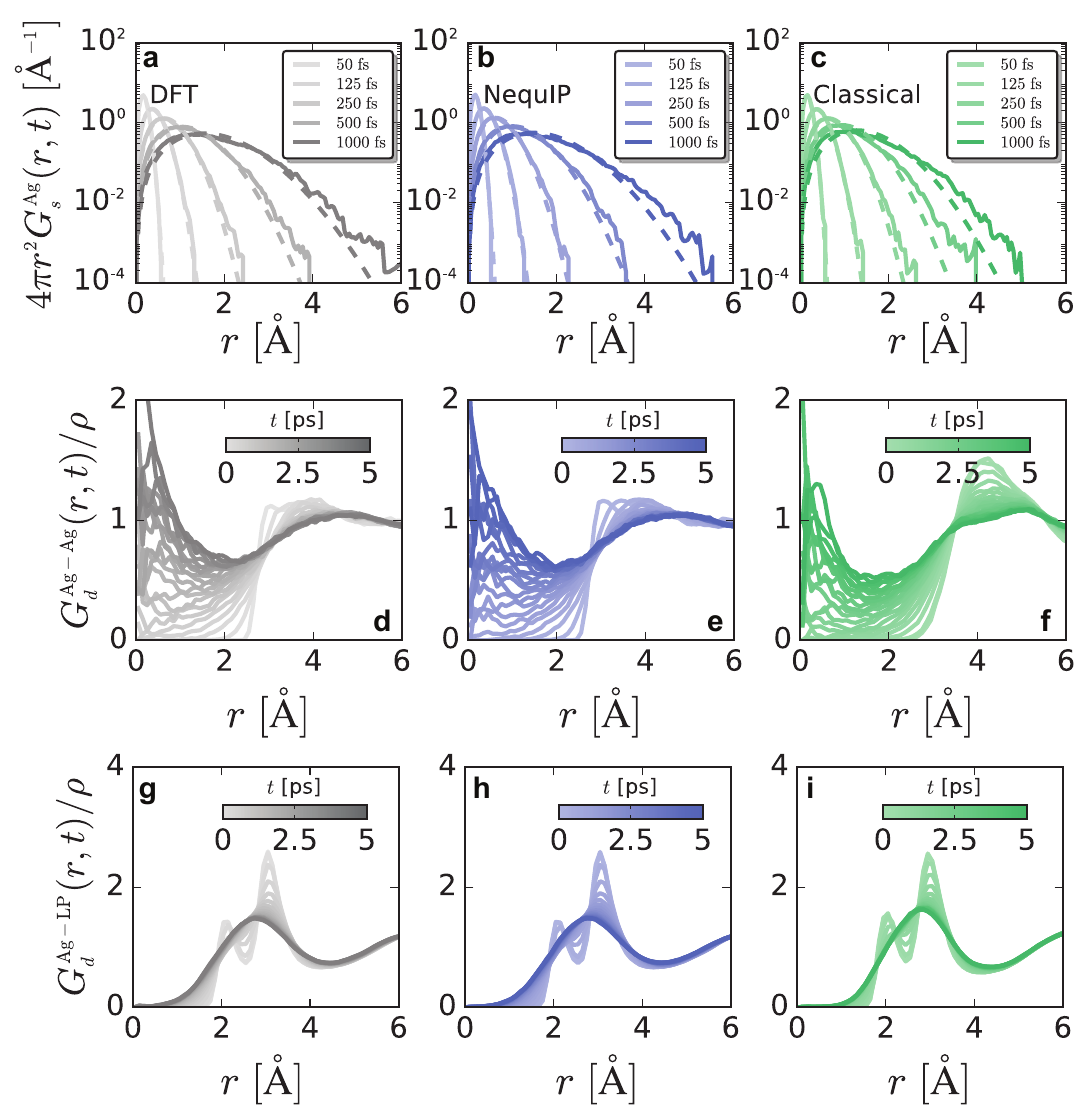}
\end{center}
\caption
{Comparison of 
(a-c) self van Hove correlation function for cation correlations,
(d-f) distinct van Hove correlation function for cation-cation pair correlations, and 
(g-i) distinct van Hove correlation function for cation-lone pair correlations 
obtained from (a,d,g) DFT, (b,e,h) MLP, and (c,f,i) FF models.
}
\label{fig:vh}
\end{figure}
%@@@@@@@@@@@@@@@@@@@@@@@@@@@@@@@@@@@@@@@@@@@@@@

\begin{figure}[tb]
\begin{center}
\includegraphics[width=0.49\textwidth]{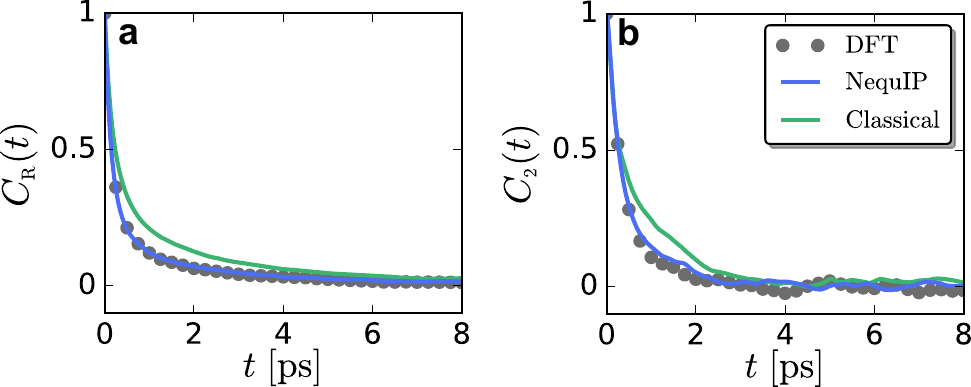}
\end{center}
\caption
{(a) Tetrahedral rotor function time correlation functions (TCFs), $C_2(t)$, characterizing I$^-$ lone pair rotational dynamics, and 
(b) cage residence correlation function, $C_{\rm R}(t)$, of Ag$^+$ in the solvation shell of I$^-$.
}
\label{fig:diff}
\end{figure}

%%
%
%%
%
We now characterize correlations in both space and time through van Hove correlation functions.
We separate the van Hove spacetime correlation functions into the self and distinct parts.
The self van Hove function, 
\begin{equation}
G_s(r,t) = \avg{ \frac{1}{N} \sum_i\delta\para{\len{\rb_i(0)-\rb_i(t)}}},
\end{equation}
quantifies spacetime correlations between a particle located at $r=0,t=0$ and itself at a position $r$ after time $t$. 
We focus specifically on $G_s(r,t)$ for Ag$^+$, because I$^-$ does not diffuse, and use it to determine whether cation translational motion occurs through smooth diffusion or discrete hopping events.
If translations occur through hopping, $G_s(r,t)$ will display peaks at well-defined distances as time is increased~\cite{Bala2020}.
If translational motion is diffusive, $G_s(r,t)$ will be smooth with only small deviations from a Gaussian for all $t$.
For all three models, we find that $G_s(r,t)$ is smooth and Gaussian-like for all $t$, suggesting that cationic translational motion is diffusive, Fig.~\ref{fig:vh}a.
Self van Hove functions of individual cations exhibit the same behavior, Fig.~S2.
However, collective dynamics involve correlated motion over large lengthscales, and the small systems accessible to DFT might not be able to fully capture dynamics on these larger scales. 
Therefore, we also modeled a significantly larger system with the FF model,
and we find peaks in the corresponding $G_s(r,t)$ indicative of collective hopping transport, Fig.~S4, at distances larger than those that can be efficiently modeled with DFT simulations.
However, the local fluctuations that give rise to a single hopping event should be relatively insensitive to system size, to a good approximation, and we focus on the small systems for which electronic structure calculations are readily accessible in the rest of this work and anticipate that our findings will hold true for larger systems as well. 
The distinct part of the van Hove function, 
\begin{equation}
G_d(r,t) = \avg{ \frac{1}{N} \sum_i \sum_{j\ne i} \delta\para{\len{\rb_i(0)-\rb_j(t)}}},
\end{equation}
measures time-dependent correlations between an ion located at the origin at time zero and other particles at $r$ at time $t$. 
At $t=0$, the distinct van Hove function is equal to the radial pair distribution function, $G_d(r,0)=g(r)$.
We focus specifically on $G_d(r,t)$ for cation-cation correlations and correlations between Ag$^+$ and iodide MLWFCs, Fig.~\ref{fig:vh}b,c.
We find that all models produce qualitatively similar distinct van Hove functions, although those produced by the FF are more structured than those from DFT and the MLP. 
The distinct van Hove function for cation correlations, $G_d^{{\rm Ag-Ag}}(r,t)$, increases at small $r$ with increasing $t$, which indicates that the cation initially at $r=0$ diffuses and is replaced by another cation on picosecond timescales, consistent with superionic behavior.
The distinct van Hove functions for cation-MLWFC correlations, $G_d^{{\rm Ag-LP}}(r,t)$, indicate that this diffusion process is accompanied by a change in MLWFC coordination. 
The initial two peaks between $r=2$~\AA~and $r=4$~\AA \ blur into a single peak on the same timescale of Ag$^+$ diffusion, suggesting that there is a coupling between cation and MLWFC dynamics.
To quantify the translational motion of a cation from one coordination environment to another, we study a time correlation function used in our previous work~\cite{Dhattarwal2024paddle} and inspired by cage correlation function analyses of supercooled liquids~\cite{rabani1997calculating}. 
To do so, we first compute the neighbor list of the $i$th cation at time $t$, $l_i(t)$. 
The neighbor list is a vector of length $N$, where each element of the vector indexes an atom in the system and equals one if the atom is in the first coordination shell and equals zero if it is not. 
The coordination shell is defined by the iodide ions within the first peak of the corresponding Ag-I $g(r)$.
Using these neighbors lists, we can compute the fraction of original neighbors that remain in the coordination cage after a time $t$,
\begin{equation}
\xi^{\rm out}_i(0,t) = \frac{ l_i(0)\cdot l_i(t)}{l_i(0)\cdot l_i(0)},
\end{equation}
such that $1-\xi^{\rm out}_i(0,t)$ is the fraction of neighbors that have \emph{left} the coordination cage of particle $i$ between $0$ and $t$.
Similarly, we define the fraction of neighbors at time $t$ that were present at time $t=0$,
\begin{equation}
\xi^{\rm in}_i(0,t) = \frac{ l_i(0)\cdot l_i(t)}{l_i(t)\cdot l_i(t)},
\end{equation}
such that $1-\xi^{\rm in}_i(0,t)$ is the fraction of neighbors that have \emph{entered} the coordination cage of particle $i$ between $0$ and $t$.
These functions measure changes in the coordination cage due to ions moving out of the cage or in to the cage, respectively.
Because we are only concerned with whether or not the cage has changed due to cation diffusion, and not by how much the identity of the cage has changed, we also define corresponding indicator functions,
\begin{equation}
h_i^{\rm out}(0,t)=\Theta\para{1-\xi^{\rm out}_i(0,t)}
\end{equation}
and
\begin{equation}
h_i^{\rm in}(0,t)=\Theta\para{1-\xi^{\rm in}_i(0,t)},
\end{equation}
where $\Theta(x)$ is the Heaviside step function.
If the number of neighbors in the silver ion's coordination cage increases between $0$ and $t$, $h_i^{\rm out}(0,t)=0$ and $h_i^{\rm in}(0,t)=1$.
If the number of neighbors in the silver ion's coordination cage decreases between $0$ and $t$, $h_i^{\rm out}(0,t)=1$ and $h_i^{\rm in}(0,t)=0$.
If the number of neighbors are the same at $0$ and $t$, then $h_i^{\rm out}(0,t)=1$ and $h_i^{\rm in}(0,t)=1$.
Finally, we use these indicator functions to compute the cage residence TCF,
\begin{equation}
C_{\rm R}(t) = \avg{h_i^{\rm out}(0,t)h_i^{\rm in}(0,t)},
\end{equation}
which will decay from 1 to 0 over some correlation time for a diffusing particle that changes its coordination cage, as one might expect for an ion diffusing within a solid.
For all three models, $C_{\rm R}(t)$ decays on similar timescales, Fig.~\ref{fig:diff}a.
We estimate the correlation times, $\tau_{\rm R}$, by integrating $C_{\rm R}(t)$ from 0 to 10~ps,
which yields $\tau_{\rm R}^{({\rm DFT})} \approx 570$~fs, $\tau_{\rm R}^{({\rm MLP})} \approx 600$~fs,
and $\tau_{\rm R}^{({\rm FF})} \approx 950$~fs.
The TCFs for the DFT and MLP models are identical, while the FF model decays slower.
This slower decay in the FF model is again due to the stiffer lattice it produces,
as well as the neglect of the impact of electronic paddlewheels on dynamics, as discussed further below.
%

%%
%
%@@@@@@@@@@@@@@@@@@@@@@@@@@@@@@@@@@@@@@@@@@@@@@
%@@@@@@@@@@@@@@@@@@@@@@@@@@@@@@@@@@@@@@@@@@@@@@

%%
%
We recently demonstrated that cation diffusion in AgI is coupled to the rotation of the iodide electron density, resulting in electronic paddlewheels~\cite{Dhattarwal2024paddle}.
To quantify the existence of electronic paddlewheels, we computed time correlation functions that quantify electronic rotations.
To characterize electronic rotations, we quantify the orientation of the tetrahedron-like unit formed by the four MLWFCs of an iodide ion using a tetrahedral rotor function, $\Mb_\lambda$, of order $l=3$, such that $\lambda$ labels the $(2l+1)$ functions for each $l$. 
Here we focus on $\lambda=2$ as in our previous work~\cite{Dhattarwal2024paddle}, and compute the time correlation function,
\begin{equation}
C_2(t) = \frac{ \avg{\Mb_2(0)\Mb_2(t)} }{ \avg{\Mb_2^2(0)}},
\end{equation}
where the rotor function is
\begin{equation}
\Mb_2(t) = \frac{3\sqrt{5}}{40} \sum_{i=1}^4 \para{5x_i^3 - 3x_i r_i^2},
\end{equation}
and $\rb_i=(x_i,y_i,z_i)$ is a unit vector along one of the four I-MLWFC bonds ($i$), and $r_i=\len{\rb_i}$. 
For the DFT-based AIMD simulations, $C_2(t)$ monotonically decays to zero on a timescale of approximately half a picosecond, Fig.~\ref{fig:diff}b,  $\tau_2^{({\rm DFT})} \approx 410$~fs, 
indicating that iodide MLWFCs exhibit orientational fluctuations on the same timescale as silver diffusion as measured by $C_{\rm R}(t)$. 
Here, we estimated the iodide MLWFC rotational times, $\tau_2$, by the integral of $C_2(t)$ from 0 to 10~ps.
The similar values of $\tau_2^{({\rm DFT})}$ and $\tau_{\rm R}^{({\rm DFT})}$ suggest that translational diffusion of Ag$^+$ is coupled to rotational motion of I$^-$ electron density --- electronic paddlewheels. 
We post-processed trajectories generated with NequIP to compute MLWFCs, and these MLWFC trajectories produce a $C_2(t)$ that is similar to that in the AIMD simulations, Fig.~\ref{fig:diff}b, with a correlation time of $\tau_2^{({\rm MLP})} \approx 530$~fs. 
NequIP learns the complicated, high-dimensional potential energy surface from AIMD simulations, including the effects of electron rotation-nuclear translation coupling that produces electronic paddlewheels~\cite{Colin2024,Dhattarwal2024paddle}. 
As a result, NequIP trajectories accurately include the effects of electronic paddlewheels and can be used to examine their behavior. 
We also post-processed trajectories generated with the classical FF model to compute MLWFCs. 
The resulting trajectories produce a $C_2(t)$ that decays on timescales slower than those in the AIMD and NequIP simulations, Fig.~\ref{fig:diff}b, $\tau_2^{({\rm FF})} \approx 780$~fs,
but this decay is still on a similar timescale to that of the corresponding $C_{\rm R}(t)$.
However, this decay is \emph{not} due to electronic paddlewheels but is instead due to translational fluctuations within the solid. 
As detailed below, the configurations generated with the FF model cannot be used to gain meaningful insights into electronic fluctuations, and the neglect of electronic paddlewheels results in significant differences in the underlying determinants of diffusion in the classical FF model.

%@@@@@@@@@@@@@@@@@@@@@@@@@@@@@@@@@@@@@@@@@@@@@@
\begin{figure}[tb]
\begin{center}
\includegraphics[width=0.49\textwidth]{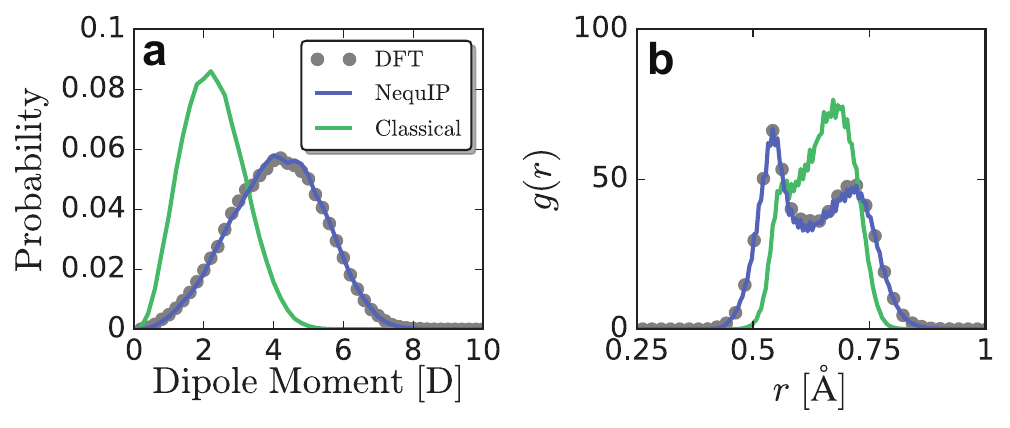}
\end{center}
\caption
{(a) Probability distribution of the dipole moment of I$^-$ ions and (b) radial distribution function of I-MLWFCs correlations obtained from DFT, MLP, and FF models.}
\label{fig:polarization}
\end{figure}
%@@@@@@@@@@@@@@@@@@@@@@@@@@@@@@@@@@@@@@@@@@@@@@

%@@@@@@@@@@@@@@@@@@@@@@@@@@@@@@@@@@@@@@@@@@@@@@
\begin{figure*}[tb]
\begin{center}
\includegraphics[width=0.99\textwidth]{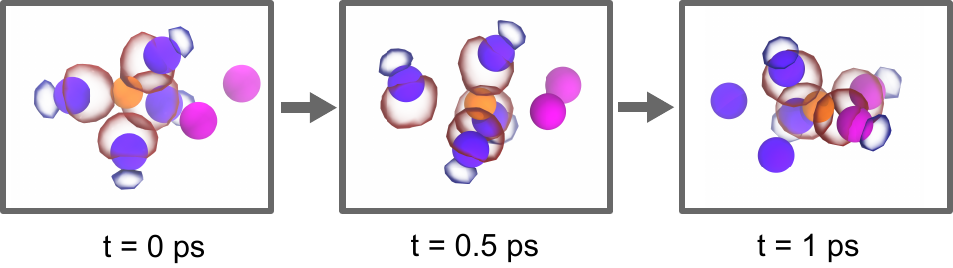}
\end{center}
\caption
{Snapshots illustrating the rotation of iodide MLWFs (surfaces) that accompany the diffusion of a silver cation (orange) as it moves from one coordination environment to another.
Iodide ions are drawn as purple spheres, with two iodides drawn in a brighter shade of purple; these two iodides initially do not coordinate silver and become part of the coordination shell in the last frame.
Select MLWFs are drawn for iodides close to the silver cation, all at an isosurface contour of 0.04~Bohr$^{-3}$.}
\label{fig:mlwfs}
\end{figure*}
%@@@@@@@@@@@@@@@@@@@@@@@@@@@@@@@@@@@@@@@@@@@@@@

\subsection{Electronic Structure Fluctuations}

We first quantify the electronic response of iodide through the distribution of its dipole moment, Fig.~\ref{fig:polarization}a.
The DFT and MLP simulations produce an average I$^-$ dipole moment of approximately $4$~D, while the FF model produces a dipole moment about half as large, approximately $2$~D. 
Moreover, the DFT and MLP models exhibit a broader distribution of dipole moments than the FF model. 
The larger dipole moment and larger dipole moment fluctuations in the DFT and MLP models suggests that the FF model does not adequately describe structural fluctuations that polarize I$^-$.
We can gain deeper insights into the origins of this difference in polarization between the DFT-based and FF models by computing
the radial distribution function (RDF) between I$^-$ and its MLWFCs, Fig.~\ref{fig:polarization}b.
In the DFT and MLP models, the I-MLWFC RDF is bimodal, which indicates that some MLWFCs are further away from the nucleus than others, consistent with a large polarization of the ion. 
The bimodal distribution of MLWFCs around I$^-$ is consistent with previous work that demonstrated that three of the four MLWFCs are stretched away from the nucleus and point toward Ag$^+$ ions~\cite{wood2006}.
Indeed, we observe that the average MLWFC coordination number around Ag$^+$ is three.
In contrast, the FF model does not produce a bimodal RDF, Fig.~\ref{fig:polarization}b.
The resulting peak in the RDF is much narrower than that produced by the DFT-based models, consistent with smaller polarization fluctuations.
The peak is asymmetric, but the RDF does not exhibit a pronounced minimum.
Although post-processing configurations generated by the FF model can capture the average tetrahedral coordination of the I$^-$ MLWFCs (evidenced by the interionic RDFs and joint probability distributions), the FF model cannot capture the polarization of I$^-$ and its resulting coupling to Ag$^+$ diffusion.
To illustrate the importance of electronic paddlewheels in the translational diffusion of Ag$^+$,
we visualized MLWFs along a portion of the trajectory where Ag$^+$ moves from one coordination environment to another, contributing to the decay of $C_{\rm R}(t)$. 
We show the four closest MLWFs coordinating the Ag$^+$ of interest at three different times along the process, $t_1=0$~ps, $t_2=0.5$~ps, and $t_3=1$~ps, Fig.~\ref{fig:mlwfs}.
We also show two nearby iodides (highlighted in a different shade of purple) that are not coordinating the Ag$^+$ at $t_1$ but are coordinating the Ag$^+$ at $t_3$. 
At time $t_1$, the MLWFs of four I$^-$ ions are tetrahedrally coordinated around the Ag$^+$ ion, which corroborates the average structure identified above.
Each MLWF has a teardrop shape that results from the directional nature of the cation-anion interactions.
Of these four iodide ions, three are coordinating the cation more strongly; the MLWF of the fourth iodide does not touch the cation at the drawn isosurface value.
In addition, the bond order between Ag$^+$ and the three strongly coordinating I$^-$ ions is 0.5, which suggests a partially covalent interaction consistent with previous work on related materials~\cite{aniya1992chemical,madden1996covalent}. 
In contrast, the fourth I$^-$ has a smaller bond order of 0.24, consistent with it coordinating the cation less strongly. 
This structural asymmetry highlights the structural anharmonicity of Ag-I interactions. 
At time $t_2$, the Ag$^+$ is transitioning between one coordination environment and another. 
The cation is moving away from one of the I$^-$ ions, which no longer has its MLWF pointed toward the cation
and exhibits a low bond order of 0.06.
The other three iodides coordinate Ag$^+$ more strongly, as evidenced by their MLWFs stretching even more towards the Ag$^+$ and their bond order increasing to approximately 0.55.
Clearly, the I$^-$ MLWFs are dynamic and both rotate and polarize as the cation diffuses through the lattice. 
At $t_3$, the Ag$^+$ has completed a transition from one coordination to another, having exchanged two iodides for two new iodides;
the old iodides are no longer drawn with their MLWF surfaces.
The new coordination structure resembles that of the initial configuration, and the cation-anion bond orders are similar as well. 
The overall MLWF trajectory clearly indicates the presence of electronic paddlewheels --- reorientation of electron density couples to diffusion.
Our analysis of electron density fluctuations emphasizes the roles of both polarization and chemical bonding in ionic diffusion and is compatible with previous theories that have separately emphasized the importance of each. 
Polarization of ions induced by their environment result in instantaneously aspherical electronic density of ions that fluctuates as ions diffuse, clearly influencing ionic dynamics~\cite{wilson1996quadrupole,salanne2011polarization,aguado2003multipoles,madden1996covalent}. 
Local fluctuations between covalent and ionic bonds have been proposed to be an important component of solid-state ion conduction~\cite{aniya1992chemical}.
Similarly, asymmetric coordination of mobile cations by anion electron density and resulting fluctuations in bonding are also relevant~\cite{wood2006,wood2021paradigms}. 
Electronic paddlewheels are compatible with and unify these concepts. 
As a silver cation moves from one coordination environment to another, it indeed starts out asymmetrically coordinated. 
However, when the cation translates toward the transition state, the orientation of the electronic structure (or MLWFs) changes, necessitating a description beyond statistical averages of electron density. 
During this process, the dipole moments of the iodides fluctuate, consistent with the importance of polarization,
but the covalency of the various silver-iodide interactions also changes, though we do not observe a complete switch from ionic to fully covalent bonds.
Therefore, electronic paddlewheels involve these previously proposed concepts, with the additional concept of dynamic orientational disorder of the local electron density that produces a coupling between ionic translations and electronic rotations.

\subsection{Electronic Paddlewheels Produce Frictional Fluctuations}

The coupling of electronic rotations to ion diffusion should lead to fluctuations in the force on a diffusing ion beyond those arising from ionic translations. 
To investigate the role of electronic paddlewheels in determining the effective forces relevant to diffusion, we work within the framework
of the generalized Langevin equation (GLE) for Brownian motion of a particle in one-dimension~\cite{boon1991molecular,zwanzig2001nonequilibrium,daldrop2019mass,daldrop2017external}:
\begin{equation}
m\ddot{x}(t) = -\int_0^\infty dt' \Gamma(t') \dot{x}(t-t') + F_{\rm R}(t),
\end{equation}
where $x(t)$ is the position of the particle at time $t$, $\Gamma(t)$ is the memory function, and $F_{\rm R}(t)$ is the random force that obeys the fluctuation-dissipation relation, $\avg{F_{\rm R}(t) F_{\rm R}(t')}=\kT \Gamma(t-t')$, and single and double dots indicate first and second derivatives with respect to time.
We focus on the GLE for a cation and we compute the memory function by discretizing and iteratively solving the integral equation~\cite{harp1970time,berne1970calculation,boon1991molecular,daldrop2019mass}
\begin{equation}
m \frac{d C_v(t)}{dt} = -\int_0^t dt' \Gamma(t-t') C_v(t').
\end{equation}
The memory function or time-dependent friction, $\Gamma(t)$, describes the influence of past time-dependent correlations on the dynamics of a particle at the current time $t$. 
In the Markovian limit of no memory, the memory function reduces to $\Gamma(t)=\gamma\delta(t)$, where $\gamma$ is the friction coefficient,
and in general $\gamma=\int_0^\infty dt \Gamma(t)=\frac{\kT}{D}$.
Before examining $\Gamma(t)$, we can first quantify the force fluctuations on a silver cation arising from its environment, which can be more physically intuitive. 
To quantify these force fluctuations, we compute the force-force time correlation function,
\begin{equation}
C_{\rm F}(t) = \avg{\mathbf{F}(0)\cdot \mathbf{F}(t)}.
\end{equation}
The force-force time correlation function is related to the memory function through
\begin{equation}
\hat{C}_{\rm F}(\omega) = \frac{\kT \hat{\Gamma(\omega)}}{\len{ m \omega^2 - i \omega \hat{\Gamma}(\omega)}^2},
\end{equation}
where $\hat{f}(\omega)$ indicates the Fourier transform of $f(t)$, and the memory function is defined to be the single-sided form equal to zero for times less than zero~\cite{daldrop2019mass,daldrop2017external}.
%

%@@@@@@@@@@@@@@@@@@@@@@@@@@@@@@@@@@@@@@@@@@@@@@
\begin{figure}[tb]
\begin{center}
\includegraphics[width=0.49\textwidth]{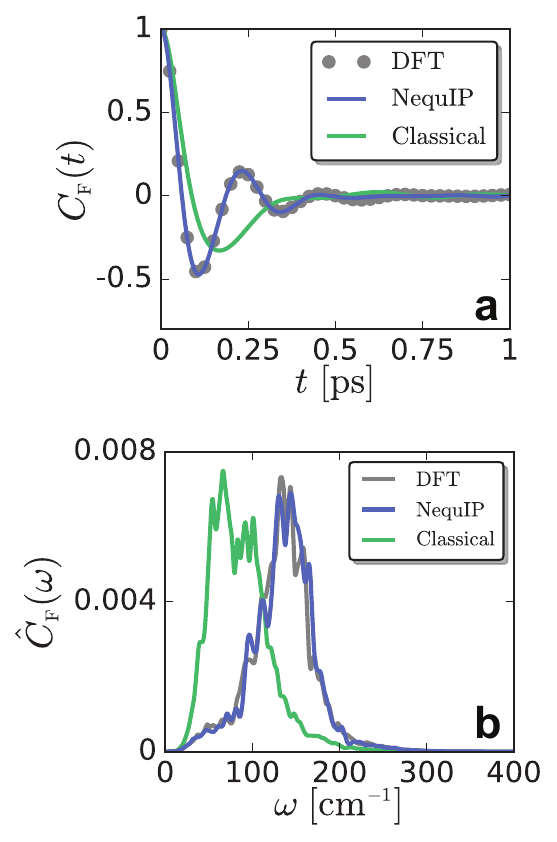}
\end{center}
\caption
{(a) Normalized force-force time correlation functions, $C_{\rm F}(t)$ for the DFT, NequIP, and Classical force field simulations, and (b) their corresponding power spectra, $\hat{C}_{\rm F}(\omega)$. }
\label{fig:cff}
\end{figure}
%@@@@@@@@@@@@@@@@@@@@@@@@@@@@@@@@@@@@@@@@@@@@@@

%@@@@@@@@@@@@@@@@@@@@@@@@@@@@@@@@@@@@@@@@@@@@@@
\begin{figure}[tb]
\begin{center}
\includegraphics[width=0.45\textwidth]{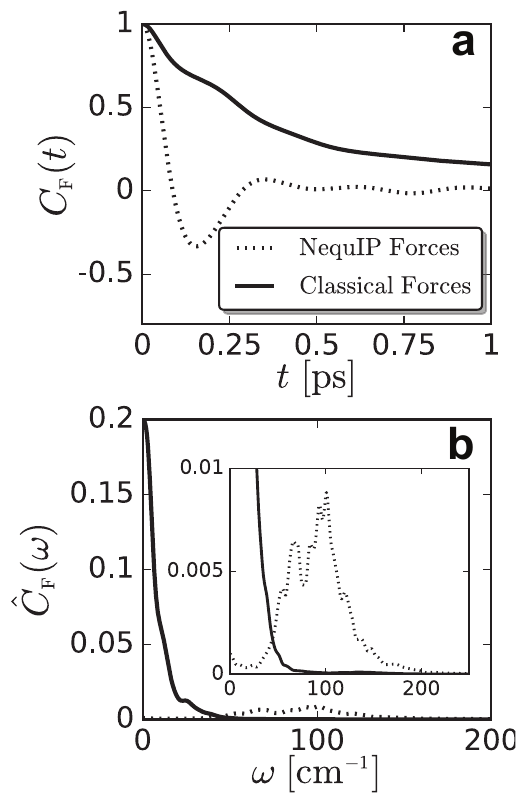}
\end{center}
\caption
{(a) Normalized force-force time correlation functions, $C_{\rm F}(t)$ obtained by post-processing classical configurations with NequIP forces and by post-processing NequIP configurations with classical forces,
as well as (b) the resulting power spectra, $\hat{C}_{\rm F}(\omega)$. 
The inset shows a zoomed in version of $\hat{C}_{\rm F}(\omega)$ to highlight the spectrum
produced by post-processing classical configurations with NequIP forces.}
\label{fig:cfpost}
\end{figure}
%@@@@@@@@@@@@@@@@@@@@@@@@@@@@@@@@@@@@@@@@@@@@@@

%%
%
The DFT and MLP trajectories produce the same $C_{\rm F}(t)$, while the correlation function differs in the FF model, Fig.~\ref{fig:cff}.
All force-force TCFs decay on timescales less than a picosecond. 
DFT and the MLP produces a $C_{\rm F}(t)$ with a faster decay and larger oscillations than the classical FF, which results in the DFT/MLP $\hat{C}_{\rm F}(\omega)$ exhibiting peaks at higher frequencies. 
It is important to note that for an unrestrained particle, $\int_0^\infty dt C_{\rm F}(t)=0$, and all three models satisfy this constraint. 
DFT and the MLP include the effects of electronic paddlewheels, while the FF model does not, such that the differences may be attributed to local electronic fluctuations. 
To examine this point further, we post-processed the MLP trajectories to compute classical FF forces and the resulting $C_{\rm F}(t)$ and $\hat{C}_{\rm F}(\omega)$, shown in Fig.~\ref{fig:cfpost}a,b.
The MLP configurations are inconsistent with the FF forces, best illustrated by the fact that $\int_0^\infty dt C_{\rm F}(t)=\hat{C}_{\rm F}(\omega=0)\approx0.2$. 
In addition, $C_{\rm F}(t)$ decays much slower than the full MLP correlation function.
Therefore, the decay of force correlations in the MLP system arises from more than the pairwise fluctuations contained in the FF model. 
In contrast, doing the opposite process, computing MLP forces from the classical FF trajectory, produces $C_{\rm F}(t)$ and $\hat{C}_{\rm F}(\omega)$ that agree reasonably well with those produced completely by the FF model. 
This agreement may be expected because the MLP can describe configurations both with and without electronic disorder, while the FF model can only describe configurations without electronic disorder. 
However, the force correlations functions still violate the constraint $\hat{C}_{\rm F}(0)=0$, albeit to a much lesser extent than the FF-processed MLP trajectories. 
Because the force-force TCFs computed in this manner violate the constraint of $\hat{C}_{\rm F}(0)=0$, we do not attempt to further analyze the dynamics that result from their corresponding GLEs. 
%

%@@@@@@@@@@@@@@@@@@@@@@@@@@@@@@@@@@@@@@@@@@@@@@
\begin{figure}[tb]
\begin{center}
\includegraphics[width=0.40\textwidth]{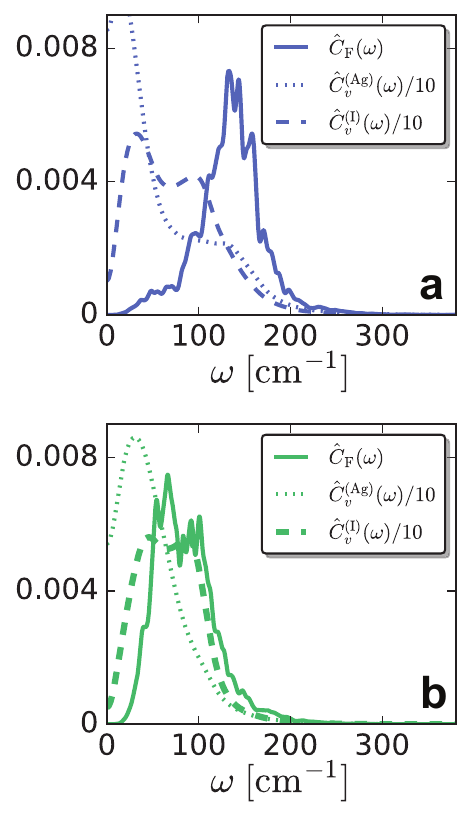}
\end{center}
\caption
{Power spectra of the force-force time correlation functions, $\hat{C}_{\rm F}(\omega)$,
compared to the corresponding velocity autocorrelation function, $\hat{C}_v(\omega)$, for (a) NequIP
and (b) classical models.}
\label{fig:fvdos}
\end{figure}
%@@@@@@@@@@@@@@@@@@@@@@@@@@@@@@@@@@@@@@@@@@@@@@

\begin{figure}[tb]
\begin{center}
\includegraphics[width=0.40\textwidth]{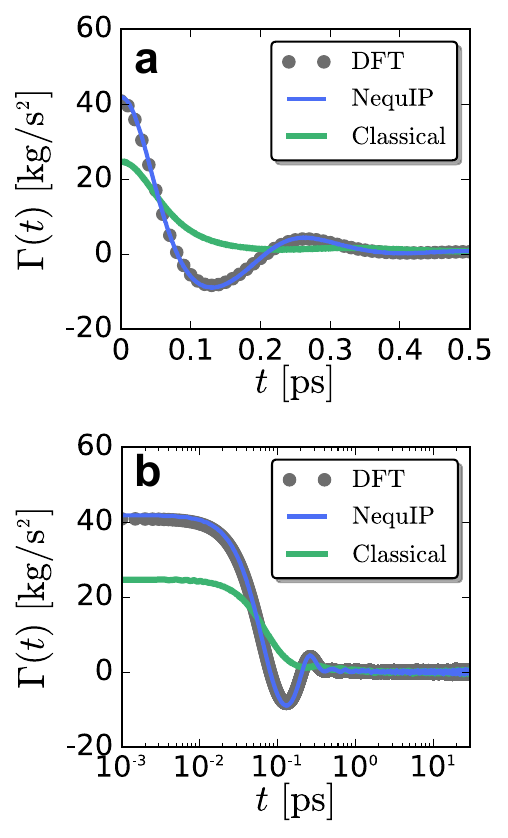}
\end{center}
\caption
{Memory function, $\Gamma(t)$, determined from DFT, NequIP, and Classical simulations.}
\label{fig:gamma}
\end{figure}

%%
%
When overlayed with the vibrational densities of states, $\hat{C}_{\rm F}(\omega)$ suggest interesting correlations between translational dynamics and forces, Fig.~\ref{fig:fvdos}.
The shape and frequencies of the $\hat{C}_{\rm F}(\omega)$ produced by DFT/MLP cannot be fully explained by the shapes of the vibrational densities of states for Ag$^+$ and I$^-$, $\hat{C}_v^{({\rm Ag})}(\omega)$ and $\hat{C}_v^{({\rm I})}(\omega)$, respectively.
The peak in the DFT/MLP $\hat{C}_{\rm F}(\omega)$ occurs at frequencies where the vibrational densities of states are low and tailing off, Fig.~\ref{fig:fvdos}a. 
In contrast, the classical FF produces a $\hat{C}_{\rm F}(\omega)$ that overlaps completely with the populated frequencies in the corresponding vibrational densities of states, Fig.~\ref{fig:fvdos}b.
The overlap between the classical force and vibrational power spectra suggest that the force fluctuations are determined by the translational fluctuations of the ions. 
In contrast, the lack of overlap between the DFT/MLP force and vibrational power spectra suggest that additional modes significantly contribute to the force fluctuations beyond translational dynamics.
We suggest that these fluctuations are indeed those due to orientational fluctuations of the electron density --- electronic paddlewheels.
We now turn our attention to the memory functions, or time-dependent friction, $\Gamma(t)$, Fig.~\ref{fig:gamma}. 
We first note that the friction coefficient is larger in the classical FF model system, consistent with its smaller diffusion coefficient; $\gamma^{({\rm FF})}=3.3\times10^{-12}$~kg/s, while $\gamma^{({\rm DFT})}=2.2\times10^{-12}$~kg/s and $\gamma^{({\rm MLP})}=2.5\times10^{-12}$~kg/s.
The initial decay of the DFT/MLP memory function is slightly faster than that of the FF memory function, consistent with smaller friction in the MLP system.
The DFT/MLP $\Gamma(t)$ exhibits oscillations at approximately the same timescale as those that appear in $C_v(t)$, the velocity autocorrelation function.
These oscillations in the memory $\Gamma(t)$ can be attributed to caging effects~\cite{boon1991molecular}.
Similar caging effects are not observed in the FF memory function, and the differences between the DFT/MLP and FF memory functions give rise to the significant differences in their respective velocity autocorrelation functions. 
The relatively simple, positive decay of the FF $\Gamma(t)$ leads to the negative decay of the corresponding Ag$^+$ $C_v(t)$ and the resulting smaller diffusion coefficient.
In contrast, the negative minimum and negative decay in the DFT/MLP $\Gamma(t)$ leads to the positive decay of the corresponding Ag$^+$ $C_v(t)$ and the resulting larger diffusion coefficient. 
Despite nearly the same pair structure, the qualitative form of the memory function differs between the DFT/MLP and FF models. 
The FF model includes only pairwise interactions and, as a result, cannot capture the many-body forces that arise from orientational disorder of local electron density, such that the time-dependent friction results from translational fluctuations only. 
The DFT and MLP models include the effects of local electronic disorder and are able to describe the time-dependent friction that results from `caging effects' on the cation by electronic rotations.
Electronic paddlewheels impact the dynamics of ion conduction by significantly changing the time-dependent friction on the diffusing ion. 
%

%@@@@@@@@@@@@@@@@@@@@@@@@@@@@@@@@@@@@@@@@@@@@@@
%@@@@@@@@@@@@@@@@@@@@@@@@@@@@@@@@@@@@@@@@@@@@@@

%%
%
Finally, we note that a measure of the duration of the memory in each system is given by
the effective memory time,
\begin{equation}
\tau_{\Gamma} = \frac{\gamma}{\Gamma(0)} = \frac{\kT}{D\Gamma(0)},
\end{equation}
which yields similar timescales for the DFT and MLP models, 
$\tau_{\Gamma}^{({\rm DFT})}\approx 53$~fs and $\tau_{\Gamma}^{({\rm MLP})}\approx 60$~fs.
However, the effective memory time is significantly longer in the FF system, $\tau_{\Gamma}^{({\rm FF})}\approx 134$~fs, consistent with the slower decay of the corresponding memory function. 
Physically, the shorter effective memory times in the DFT and MLP models originate from the additional fluctuations provided by local electronic fluctuations.
The FF model lacks the additional many-body forces resulting from electronic paddlewheels and as a result can only lose memory via fluctuations in pairwise distances. 

\section{Conclusions}

It is becoming increasingly clear that hidden disorder within electronic degrees of freedom plays an important role in many materials~\cite{Colin2024,aeppli2020hidden}, like electronic paddlewheels in solid-state ion conductors~\cite{Dhattarwal2024paddle}.
Modeling these local fluctuations in electronic structure necessitates a description of quantum mechanics at some level, but this greatly increases the computational cost of modeling these systems.
Here, we demonstrate that machine learning-based neural network potentials can capture the underlying physics of ion conduction sufficiently well to use the configurations they generate to predict electronic properties within the Born-Oppenheimer approximation, including local electronic fluctuations. 
We suggest that these much more efficient models of the potential energy surface can be used to rapidly sample configurations with high accuracy, and then electronic properties can be computed in post-processing, which can be sped up by performing calculations on different configurations in parallel. 
Ideally, this second, more intense step could also be replaced with machine learning models that predict electronic properties given a nuclear configuration to avoid the cost of electronic structure calculations altogether~\cite{Gao2022SCFNN,Dhattarwal2023SCFNN}. 
This general approach can enable the exploration of electronic effects on large length and time scales, unveiling new and important insights into materials.
We also used these models to demonstrate the effects of electronic paddlewheels on the dynamics of ion conduction through a generalized Langevin equation-based analysis. 
The classical force field model is unable to capture the many-body interactions that result from electronic paddlewheels and, as a result, frictional forces and the resulting dynamics are determined solely by translational fluctuations of the ions. 
The neural network potential and the DFT simulations include the effects of electronic paddlewheels and produce frictional forces that cannot be explained by translational dynamics alone. 
Instead, the rotational fluctuations of iodide local electron density create additional frictional forces on the silver cations, and the resulting time-dependent friction produces faster diffusion than observed in the classical force field model.
The impact of electronic paddlewheels on friction suggests that ion dynamics could be controlled by tuning local electronic structure fluctuations in solid-state ion conductors. 
%

%**********************************************************************************************************%
\section{Acknowledgements}
This work was supported by the U.S. Army DEVCOM ARL Army Research Office (ARO) Energy Sciences Competency, Advanced Energy Materials Program award \# W911NF-24-1-0200. The views and conclusions contained in this document are those of the authors and should not be interpreted as representing the official policies, either expressed or implied, of the U.S. Army or the U.S. Government.
We acknowledge the Office of Advanced Research Computing (OARC) at Rutgers, The State University of New Jersey for providing access to the Amarel cluster and associated research computing resources that have contributed to the results reported here.

%**********************************************************************************************************%

\bibliography{AgI-references}

\end{document}